\documentclass[12pt]{emulateapj}
\def\Hdel{H$\delta_{F}$ }
\def\Zq{z_{quench}}

\begin{document}
\title{Population synthesis models for late build-up of the red sequence}
\author{Justin J. Harker\altaffilmark{1}, Ricardo
  P. Schiavon\altaffilmark{1,2}, Benjamin J. Weiner\altaffilmark{1,3},
  S. M. Faber\altaffilmark{1}}
\altaffiltext{1}{UCO/Lick Observatory, Dept.\ of Astronomy \& Astrophysics,
University of California, Santa Cruz, CA 95064}
\altaffiltext{2}{Department of Astronomy, University of Virginia, P.O. Box
  3818, Charlottesville, VA 22903-0818}
\altaffiltext{3}{Department of Astronomy, University of Maryland, College Park, MD 20742}
\email{jharker@ucolick.org, rps7v@astro.Virginia.EDU, bjw@ucolick.org, faber@ucolick.org}

\begin{abstract}
We present population synthesis models designed to
represent the star formation histories of $L^*$ red sequence galaxies 
(RSGs). Earlier work has shown that single-burst stellar populations
(SSPs) are unable to match Balmer line strengths simultaneously at
high and low redshift. We therefore consider alternative star
formation histories in which RSGs contain intermediate-aged stars
even at late epochs. The models are compared to Balmer \Hdel
absorption strengths, $U \! - \! B$ color data, and the number density
of red sequence galaxies from $z=1$ to $z=0$. 
We find that quenched models, models of constant star formation
histories truncated at regularly-spaced intervals, average to an RSG
population that matches the data well, showing slow evolution in color
and Balmer line strength and a rise in number density by a factor of a
few after $z=1$. The data are best fit by a turn-on of quenching at
redshifts $z=1.5-2$.
\end{abstract}

\maketitle

\section{Introduction}
\label{Intro}
Data from low redshift surveys reveal the presence of a bimodal
distribution of galaxy colors separating red sequence galaxies (RSGs)
from a diverse population of blue galaxies \citep[e.g.,][]{Str01,
Kau03}. Data from high redshift surveys such as DEEP1, DEEP2,
and COMBO-17 \citep[]{Im_00, Wei05, Bel04}
confirm that this bimodal distribution was in place by $z \simeq
1$. From their red colors, it can be
inferred that RSGs must be comprised at all redshifts largely of
old stars \citep[]{Wei05}. However, there is evidence
for a range in RSG ages at low redshift, both in field and cluster
galaxies \citep[e.g.,][]{Jor99, Tra00a, Tra01, Kun01}. Mean Balmer
line equivalent widths (EWs) in many nearby RSGs are high
enough to indicate 
the presence of an intermediate-age component, implying that the data
are not consistent with a red sequence that was fully in place and
passively evolving after $z=1$ \citep[]{SchL06}.

To extend modeling of the red sequence, this Letter considers two
other types of star formation histories that have recently been 
mentioned in the literature. Trager et al.~(2000b) introduced
``frosting'' models, in which an early burst of star formation is
followed by a later small burst. These models allow for a
range in inferred mean stellar age even for a fixed epoch of star
formation onset. Gebhardt et al.~(2003) developed a second type
of frosting model characterized by continuous, exponentially declining
star formation following a large initial burst, utilizing these models
for their ability to fade significantly (roughly 1.5 to 2 magnitudes
in $M_{B}$) while maintaining low evolution in $U \! - \! B$ color.
We adopt the exponential frosting models as we wish to
maintain roughly constant red colors while enhancing \Hdel at late
times.

The second family of models involves continuous star formation
histories halted at prescribed epochs. These are the ``quenched''
models proposed by Bell et al.~(2004). Both COMBO-17 \citep[]{Bel04}
and DEEP2 \citep[]{Wil06, Fab06} found a rise in the number of
RSGs after $z\sim 1$. These authors suggest that some process must 
be turning blue galaxies into RSGs at late epochs. Possibilities
include spiral-spiral mergers \citep[e.g.,][]{Mih96, Cox06}, AGN or
other feedback processes \citep[e.g.,][]{Sil05, Cro06}, or scenarios
that combine the two \citep[e.g.,][]{Spr05a, Hop05}.

As formulated here, SSP models and exponential frosting models
assemble completely and form virtually all of their stellar mass well
before $z \sim 1$. The number of galaxies on the red
sequence using these models therefore remains essentially
constant after this epoch (barring mergers, which could reduce number
density). Quenching models are fundamentally different in that they
add new galaxies to the red sequence even at late epochs and therefore
allow for the number of RSGs to rise with time. They furthermore allow
for a red sequence in which mean spectral and photometric properties
may be strongly influenced by recent arrivals at late times.

The observed late rise in the number density of RSGs therefore
naturally favors quenching models, either ones
in which an early constant star formation phase is simply turned off
via feedback, or ones in which the quenching phase is preceded by a
burst of formation of young stars, as occurs in many merging-galaxy
models. We develop both types of quenched models, and also
develop the aforementioned frosting models to test formation scenarios
against our observational constraints in $U \! - \! B$ color and \Hdel
EW. The older SSP models are retained, even though they
do not fit, as a comparison to earlier literature. We also test the
quenched models against RSG number density as a function of redshift. 

Colors are on the Vega magnitude system, and we use a cosmology
of $\Omega_{\Lambda} = 0.7$, $\Omega_{M} = 0.3$, $h_{0}=70$.

\section{Data}
\label{Data}
Our high-redshift Balmer absorption-line data are derived from stacked
DEEP2 spectra (four redshift bins, $\sim \! 200$ galaxies/bin), while
low-redshift values have been computed in the same way from stacked
Sloan Digital Sky Survey (SDSS) spectra \citep[two mass bins, $\sim \! 5000$
galaxies/bin;][]{Eis03}. High redshift $U \! - \! B$ colors are
derived from CFHT {\it BRI} data \citep[]{Wil06},  while low redshift
$U \! - \! B$ values are taken from Weiner et al.~(2005), which are in
turn taken from the RC3 catalog \citep[]{RC3}. When selecting red
galaxies, we adopt an evolving color cut using Weiner et al.~(2005)
data, held constant at $U \! - \! B > 0.15$ prior to $z=1$ and rising
to $U \! - \! B > 0.35$ by $z=0$, assuming a linear rise in $U \! - \!
B$ as a function of redshift. The Balmer line data are described in
Schiavon et al.~(2006).

The number density of galaxies on the red sequence is taken from Faber
et al.~(2006), who use high-redshift data from DEEP2
and COMBO-17 and low-redshift data from SDSS and 2 Degree Field Galaxy
Redshift Survey data \citep[]{Nor02}.

\section{Models}
\label{Models}
The high-resolution spectral synthesis code of Bruzual \& Charlot
(2003) was used to construct the model star formation histories. All
models are developed with a Salpeter IMF \citep[]{Sal} using solar
metallicity and solar abundance ratios. 

The frosting models consist of a 1-Gyr-long burst starting at $z=5$
that consumes 80-99\% of the available gas, followed by an
exponentially-declining phase defined by an e-folding time
between 1-10 Gyr. The quenched models also begin at $z=5$,
with constant star formation rates. A series of models is
quenched starting by either redshift $z_{q} \sim 2$ or $z_{q} \sim 
1.5$, with successive models quenched uniformly in time at intervals 
of 250 Myr. (The first quenching event is timed to be slightly earlier
so that the first galaxies pass the color cut by $z_{q}$.) 
A first series of quenched models continues the constant
star-formation rate until quenching. However, to model the burst of
star formation associated with gas-rich mergers, we also develop
models that quench immediately after a 1 Gyr burst of 5x enhanced star
formation. Burst parameters are based roughly on those of moderate
strength bursts in Cox et  al.~(2006) and Springel et
al.~(2005b). These two series are referred to as ``pure'' and
``burst''  quenched models, respectively. The chosen model
formulations are of an extremely simple nature, designed to minimize
the number of input parameters. However, even these simple models
match the data reasonably well.

\begin{figure}[t]
\epsscale{1.0}
\plotone{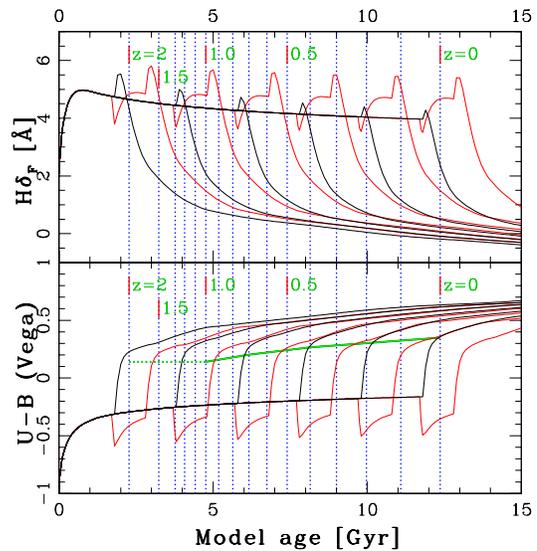}
\caption{
Behavior of quenched models in \Hdel line strength (top) and $U \! -
\! B$ color (bottom) versus time. Pure quenched models are shown as
black lines, burst-quenched models are shown in red. The green line in
the lower panel represents the cut in $U \! - \! B$ color taken from
Weiner et al.~(2005). For clarity, only every eighth model is plotted,
with an offset between the pure and burst quenched models. Redshifts
are marked by blue dotted lines in increments of 0.1 in redshift up to
$z=1.5$. The first model undergoes quenching just prior to $z=2$ in
order to pass the color cut by $z=2$. O and B stars
contribute proportionally more to the continuum than to Balmer line
strength, which causes a temporary reduction in \Hdel line strength at
the onset of the burst for the burst quenched models. Decay of those
stars also accounts for the temporary rise in \Hdel following
quenching for both burst and pure models, as the A type stars come to
dominate \Hdel after the continuum-producing high mass stars
die off.
\label{quench-hist}
}
\end{figure}

Figure~\ref{quench-hist} shows the behavior of both sets of
quenched models in EW and color. Upon quenching, all models
approach a uniform locus in $U \! - \! B$ color, 
crossing our color cut in $\lesssim$ 1 Gyr, a generic feature of
models that undergo rapid quenching. Since we are interested in
the average behavior of quenched models on the red sequence at a
given location and epoch, we take the {\it number-weighted average} of
color or \Hdel EW for those quenched models that
pass the color cut at that epoch. This amounts to assuming that
galaxies arrive at that red sequence location at a uniform rate, which
we take to be approximately that of a red $L^{*}$ galaxy. Since all
models fade to the same low EW within 2 
Gyr, at late epochs any \Hdel enhancement in the averaged population
will be dominated by the most recently quenched models.

To calculate $\phi^{*}$ for the quenched models, we assume
that the number of objects that have quenched prior to a given epoch
directly traces RSG number density. Given that successive models
quench at constant intervals in time, they therefore exhibit a linear
rise in $\phi^{*}$ with time. We normalize the number of quenched
models that have passed the color cut at $z=0.5$ to match $\phi^{*}$ at
that redshift from Faber et al.~(2006).

\section{Color and Equivalent width}
\label{Results1}

\subsection{Frosting Models}

\begin{figure*}[t]
\centerline{
\scalebox{.45}{\includegraphics[angle=-90]{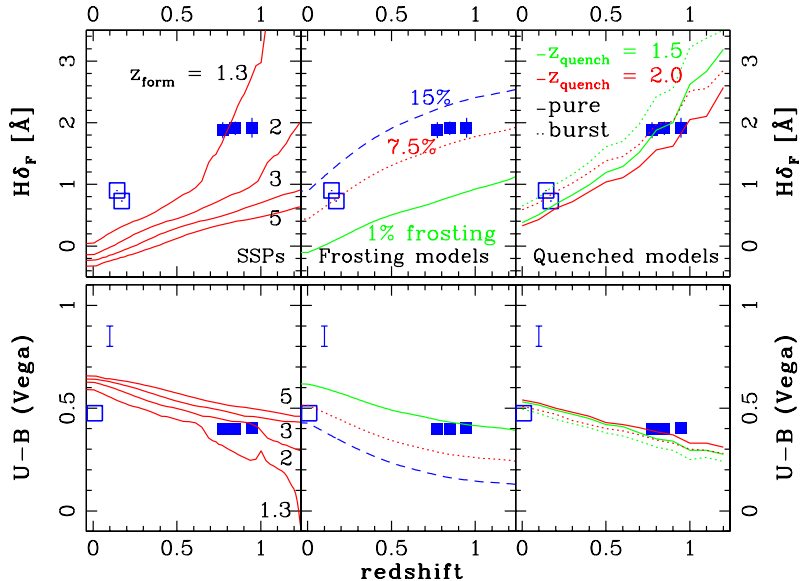}}}
\caption{
Measurements of mean \Hdel line strength (upper panels) and mean $U \!
- \! B$ (lower panels) vs. redshift. A selection of frosting
models with varying levels of ongoing star formation falling with an
e-folding time of 4 Gyr are plotted in the central panels. The pure
(solid lines) and burst (dotted lines) quenched models are plotted in
the rightmost panels, with those models that began quenching at $z=1.5$
in green and those that began quenching at $z=2.0$ in red. For
comparison, SSPs with varying epochs of formation are plotted in the
leftmost panels. Blue points represent data: low-redshift \Hdel
measurements of the highest-two mass bins for galaxies in
SDSS \citep[]{Eis03} and $U \! - \! B$ measurements
from the RC3 catalog are plotted as open squares; high-redshift \Hdel
and $U \! - \! B$ from DEEP2 \citep[]{SchL06} are plotted as open
squares. An error bar is plotted in the $U \! - \! B$ panels to
represent the mean RMS scatter in RSG $U \! - \! B$ color, which is
constant for redshifts $z<1$ \citep[]{Wei05}. 
\label{lines-colors}
}
\end{figure*}

Figure~\ref{lines-colors} shows measurements of mean \Hdel and $U \! -
\! B$ vs.~redshift for data and models. The middle panels show a
selection of individual frosting models, the different lines 
represent varying amounts of total star formation. The right panels
show both sets of averaged quenched models for two assumed epochs of
quenching onset. For comparison, SSPs with varying epochs of
formation are plotted on the left.
The SSP model that most closely matches high-$z$ \Hdel data
requires an unrealistically low epoch of formation, and furthermore
predicts very blue colors prior to $z\sim 1$. No such model has the
strong intermediate-aged stellar component at all redshifts required
to match the enhanced Balmer line EWs, as noted by Schiavon et
al.~(2006). 

Frosting models match \Hdel better due to ongoing star
formation. However, the best-fit frosting model evolves too steeply in
color. Furthermore, since the percent mass reserved
for the frosting phase affects primarily the
zeropoint of the model behavior in $U \! - \! B$ and \Hdel, to get a
better fit to color data it is necessary to tune the e-folding
time. However, doing so necessarily worsens the \Hdel fit. A single
frosting model cannot be tuned to fit simultaneously the color and
EW data at all epochs, due to the fact that ongoing star
formation means the \Hdel producing intermediate-age stars also exist
alongside a young stellar population that strongly affects $U \! - \!
B$ color.

The poor match to color data is perhaps a weak argument to rule out
frosting models since there is as yet no firm consensus on the amount
of $U \! - \! B$ color evolution on the red sequence ({\it cf.} Bell
et al.~2004 vs. Weiner et al.~2005). It may be that combinations of
frosting models are a better match to the data, but we have not yet
found any recipe that works using two-component models. Moreover,
exponential frosting models as parameterized here have two other
significant flaws. The first is their aforementioned failure to
reproduce the rise in $\phi^{*}$ RSGs since $z=1$. Second, exponential
frosting models are  forming stars at all epochs. At $z=0$, the
best-fit frosting model forms stars at a rate of $\sim 0.12
M_{\odot}/yr$ for a $10^{11} M_{\odot}$ galaxy. This level of star
formation produces $H\alpha$ emission that is comparable to the
values seen in SDSS RSG spectra \citep[]{Yan06}. However, $H\alpha$ in
these sources is thought to be excited by AGN or other hard UV
excitation sources rather than by star formation
\citep[e.g.,][]{Kau05}.

\subsection{Quenched models}

Since Balmer lines are strongest in intermediate-age stars, it 
can be seen that the SSP model fails since it has enhanced \Hdel
for only a few Gyr and can therefore not match high \Hdel over the
wide range of epochs seen. Frosting models do possess an
intermediate-aged stellar population at all redshifts, but ongoing
star formation means that  their line strengths are always diluted by
continuum from high-mass stars which also produce bluer $U \! - \! B$
colors. To fit high \Hdel at high $z$ then requires even higher star
formation rates, which produce further bluer colors, and ultimately
neither SSPs nor frosting models can fit both sets of data
simultaneously.

In contrast, both pure- and burst-quenched models fit the data rather
well. Since the averaged populations include only objects with
$U \! - \! B$ colors that pass our color cut, we obtain a good
fit to color almost by construction. By changing the epoch at which
quenching begins, we can then fit to \Hdel without disturbing
the color fit.

If the onset of quenching is assumed to begin at $z=1.5$ (green
lines), the predicted decline of \Hdel is steep, all but ruling out
the burst-quenched models although the pure-quenched models (red
lines) are still a good fit. For a quenching onset of $z=2.0$, both
types of models match the data. Without
higher-redshift \Hdel measurements, we cannot constrain the onset of
quenching very precisely, any epoch before $z=1.5$ being acceptable.

\section{Evolution of red-sequence galaxy number density}
\label{Results2}
\begin{figure}[t]
\epsscale{1.0}
\plotone{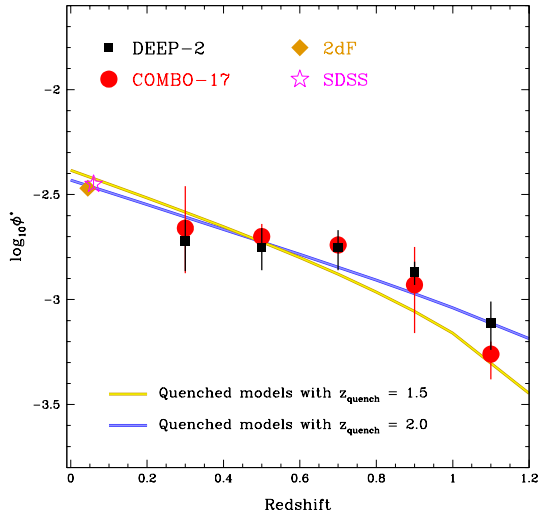}
\caption{
Rise toward low redshift of observed RSG number density compared to 
quenched models. Points represent data from Faber et al.~(2006).
Galaxies quench at a constant rate. The curves show results for the
two choices of quenching onset, with $\Zq=1.5$ in gold and $\Zq=2.0$
in blue.
\label{phi-star}
}
\end{figure}

Number density data provide an independent test of models. 
The SSP and frosting models do not evolve with time, but the quenched
model population builds up over time and so can be tested against
measured number density evolution. Quantitative results are plotted in
Figure~\ref{phi-star}, which shows model versus observed values of
$\phi^{*}$, with model $\phi^{*}$ values normalized to match
observations at $z=0.5$ as described in \S\ref{Models}. 

For either choice of the onset of quenching, the fits to data are 
within the range of observed parameters. In raw numbers, Faber et
al.~(2006) measured a rise in RSG $\phi^{*}$ by a factor of $3.6\pm0.9$
from $z=1$ to the present day. The quenched model $\phi^{*}$ rises over 
the same period by a
factor of 3.6 for $z_q=2.0$, and by a factor of 5.4 for $z_q=1.5$.
Both values are within range of the observed values, though
$z_q=2.0$ is slightly preferred, especially given the steep falloff 
above $z=1$ for the $z_q=1.5$ curve.

\section{Issues and Conclusions}
\label{Issues}
The models here are based on very simple
assumptions and merit a number of caveats. We
assume solar metallicity and solar abundance ratios at all epochs.
Higher-metallicity populations would have reduced Balmer line EWs,
and thus an average population of galaxies with evolving metallicity
would show a different trend in \Hdel strength versus time.
We also do not consider the possibility that galaxies migrate to the
red sequence at different times for different masses. Recent data
\citep[e.g.,][]{Tre05, Bun05} may indicate that the highest-mass RSGs
are fully formed by $z=1$; our model would apply to slightly dimmer
$L^{*}$ galaxies, where the growth in numbers is clearer.
Our quenched models also assume constant star formation before
quenching, whereas realistic progenitors are likely to be massive blue
galaxies that have already experienced a fall in total star formation
rate before $z=1$. A next series of models will take this into
account. Finally, our assumed uniform rate of quenching is simplistic in
the extreme. The actual rate may vary with time, as is suggested by
a possible non-linear trend in Figure~\ref{phi-star}. 

Having adopted a simple set of quenched models with
constant star formation rates truncated at a constant rate in time, we
find that these simple models do in fact match simultaneously RSG
color, line strength and galaxy number density data from the
literature. Quenched models perform better in all of these tests than
either SSPs or single frosting models. The models are not yet
sophisticated enough
to definitively predict the intial epoch of quenching. However, based
on our fits, we expect to see the red sequence come to life sometime
before $z=1.5 - 2.0$ and await future observational data that may further
clarify this issue.\\

The authors wish to thank the referee for providing valuable
feeback. This project makes use of data taken by the DEEP2 survey,
work made possible by support from National Science Foundation grants
AST 00-71198 and 00-71408. JJH acknowledges additional support from
NSF grant AST-0507483 and NASA STScI grant HST-AR-09937.01-A.

\end{document}